\documentclass[conference]{IEEEtran}
\IEEEoverridecommandlockouts

\usepackage{amsmath,amssymb,amsfonts}
\usepackage{algorithmic}
\usepackage{graphicx}
\usepackage{textcomp}
\usepackage{xcolor}
\def\BibTeX{{\rm B\kern-.05em{\sc i\kern-.025em b}\kern-.08em
    T\kern-.1667em\lower.7ex\hbox{E}\kern-.125emX}}
\usepackage{cite}
\begin{document}

\title{Towards Scalable Cross-Chain Messaging\\
}

\author{\IEEEauthorblockN{João Otávio Chervinski\textsuperscript{1}, Diego Kreutz\textsuperscript{1,2}, and Jiangshan Yu\textsuperscript{1}}
\IEEEauthorblockA{
\textsuperscript{1}Monash University, Australia \\
\texttt{\{joao.massarichervinski, jiangshan.yu\}@monash.edu} \\
\textsuperscript{2}Federal University of Pampa, Brazil \\
\texttt{diegokreutz@unipampa.edu.br}}}

\maketitle
\thispagestyle{plain}
\pagestyle{plain}

\begin{abstract}
Blockchains were originally designed as closed execution environments and lack the ability to communicate directly with external systems. To overcome this limitation, many blockchains employ relayers, external applications capable of transporting data between different blockchains. Typically, the process of relaying data is permissionless and multiple independent relayers work concurrently to transport the same information between two blockchains. While this model increases the reliability of data delivery by providing redundancy, it also introduces challenges that have not been previously discussed. In this work, we bridge this gap by discussing the shortcomings of permissionless cross-chain relaying systems and identifying three issues that adversely impact their performance, scalability and security. We take the first step towards addressing issues that hinder performance and scalability by proposing a novel protocol to enable coordination among independent relayers. Additionally, we provide an in-depth discussion about the trade-offs associated with the design of relayer coordination protocols for permissionless settings. Through this work we provide a foundation for improving cross-chain relaying services.
\end{abstract}


\section{Introduction}\label{sec:introduction}
Blockchains were originally designed as self-contained execution environments. Through research and development efforts from both industry and academia, the technology evolved rapidly and numerous blockchain platforms and applications were created. Those efforts, however, were uncoordinated, resulting in a fragmented landscape characterized by isolated and fundamentally different systems~\cite{belchior2021survey}. 
As a result, connecting blockchains requires the integration of systems built on different architectures, protocols and security guarantees. To address those challenges, constructions such as Atomic-Cross Chain Swaps~\cite{herlihy2018atomic, han2019optionality}, sidechains~\cite{kiayias2020powsidechains, back2014enabling} and notary schemes~\cite{buterin2016chain} were proposed to coordinate operations and transfer data between blockchains, a concept known as cross-chain communication.

Cross-chain communication focuses on addressing two main problems.
The first is transporting data between different blockchain systems. The second is verifying the authenticity of incoming data. Because blockchains are unable to communicate directly they require the assistance of external parties such as oracle services~\cite{zhang2016towncrier,adler2018astraea,breidenbach2021chainlink}, relayers~\cite{ibcgorelayer, hermesrelayer, yuirelayer, darwiniarelayers} or the users participating in cross-chain communication~\cite{herlihy2018atomic, deshpande2020privacy}. To validate incoming data, blockchains leverage state digests~\cite{abebe2021verifiable, zhang2020tick}, Non-Interactive Proofs of Proof-of-Work (NIPoPoWs)~\cite{kiayias2020nipopow}, Zero-Knowledge Proofs (ZKPs)~\cite{xie2022zkbridge} and Merkle Proofs~\cite{frauenthaler2020leveraging}. Those mechanisms condense the system state into proofs that are compact and verifiable.
Despite the abundance of existing strategies, there are still opportunities for the improvement of cross-chain communication, particularly when it comes to identifying and addressing the performance limitations of existing protocols~\cite{belchior2021survey}. For example, recent work highlights unadressed performance and scalability issues that affect cross-chain relayers and the transport of data in the context of the Inter-Blockchain Communication Protocol (IBC)~\cite{chervinski2023analyzing}. 

In this work we explore this research gap and focus on improving the transport of data between blockchains. In particular, we look at the shortcomings of cross-chain relayers and how they can adversely impact a system's performance, scalability and security. Relayers are an essential part of the infrastructure of various cross-chain protocols and bridging services, including the IBC protocol, Nomad Protocol, Gelato Network, Keep3r Network, Darwinia Network and LayerZero.

We identify three issues that impact cross-chain relayers deployed in a permissionless setting, namely race conditions, hindered scalability, and transaction reordering and censorship. Specifically, we discuss how those issues arise due to the lack of coordination among relayers, a common characteristic among current systems. 
We take the first step towards addressing race conditions and scalability issues by proposing a protocol to enable coordination among independent relayers. 

We design our protocol to be adaptable and operate without requiring trusted third parties. Leveraging it as a foundation, we provide an in-depth discussion about the considerations and the trade-offs associated to the design of relayer coordination protocols including incentives, task allocation and reliability. Through this work we provide a foundation for further research on cross-chain communication. Our contributions are summarized as follows:

\begin{itemize}
\item \textit{Contribution 1 (Section II)}: We identify three issues derived from the lack of coordination mechanisms in cross-chain relaying systems, namely race conditions, hindered scalability, and transaction reordering and censorship. 

\item \textit{Contribution 2 (Sections III-V)}: We introduce a novel protocol that addresses race conditions and scalability issues by enabling coordination among independent relayers.

\item \textit{Contribution 3 (Section VI)}: We provide an in-depth discussion on the trade-offs and challenges associated with the coordination of independent relayers in a permissionless setting.

\end{itemize}

\section{Cross-chain message relaying}\label{sec:relayers_background}

Relayers are off-chain applications that perform the transport of data between blockchains. They usually operate by scanning blockchains for pending cross-chain transactions or by receiving jobs from a component that aggregates and assigns data delivery tasks. 

In the context of permissionless blockchains, it is desirable for relayers to operate in a trustless fashion. This allows them to be employed without the introduction of additional points of trust. In practice, this is achieved by performing the validation of all cross-chain data in the communicating blockchains. This approach enables blockchains to detect when data has been tampered with during transport and in turn, allows the relaying process to be permissionless. This virtually removes the barrier to entry and allows anyone to deploy a relayer at any moment.

A permissionless relaying model contributes to the decentralization of the data transportation process, however, it simultaneously introduces challenges related to performance, incentive mechanisms and security. Although some blockchain services acknowledged and sought to address part of those issues~\cite{darwiniarelayers, bnbrelayerincentives, gelatocoordination}, details regarding the proposed approaches are either unavailable or lacking in depth.

\subsection{Relayer issues and games}
By leveraging a cross-chain performance evaluation tool~\cite{chervinski2023analyzing}, discussions from recent works~\cite{olshansky2023relay, ibcmovingincentives}, and conducting further investigation, we have observed three issues that negatively impact permissionless relaying systems.

\textbf{Race conditions.} In permissionless settings, such as open blockchain networks, relayers are able to join or leave the network at will. One such example is cross-chain channels established using the IBC protocol. In this case, the set of relayers working to deliver data through a channel is unknown, this is true both for the blockchains and the relayers themselves. When multiple data delivery tasks are available, the rational choice for an honest relayer is to attempt to complete as many of these as possible. This is especially true if the delivery process is incentivized by rewards. This behavior is motivated by the two types of expenses that are incurred for running a relayer~\cite{relayerfeesdata}. The first expense stems from maintaining the infrastructure required to deploy and run blockchain nodes. The second is associated with the fees that must be paid for the submission of transactions to the blockchains. While this approach seems logical from an individual relayer's standpoint, in the presence of multiple independent relayers it leads to race conditions and winner-takes-all games.

Assume two independent blockchains, $Chain_{A}$ and $Chain_{B}$, connected by a set of independent relayers $\{R_{1}, R_{2}, R_{3}\}$. Consider a pool of pending tasks containing three cross-chain transactions, $\mathcal{P}=\{tx_{1}, tx_{2}, tx_{3}\}$, to be delivered from $Chain_{A}$ to $Chain_{B}$. To deliver cross-chain transactions, relayers must pay a transaction processing fee, referred to as \textit{gas}, to block miners in the receiving blockchain. Transactions require a minimum amount of gas to be accepted, proportional to the computational effort they require to be executed and included in the blockchain by the miners. Gas fees can be adjusted and any excess amount is awarded as a bonus to miners. This means that relayers can choose to pay increased fees for the transactions they deliver, causing any rational, profit-seeking miners to prioritize the inclusion of those transaction in the blockchain. Different relayers can submit the same transaction, however, only the first valid submission is rewarded. Moreover, redundant submissions are rejected, but they still have to pay fees.
This leads to several scenarios with different outcomes:

\begin{itemize}
\item \textit{Scenario I -} Each of the relayers, $R_{1}, R_{2}$ and $R_{3}$, attempts to deliver $tx_{1}, tx_{2}, tx_{3}$ to $Chain_{B}$ as soon as they are included in $Chain_{A}$ and become available in the pool of pending tasks. All relayers use a default fee estimation algorithm and pay the same gas fee for each of the three transactions. The three relayers manage to submit all three transactions each to $Chain_{B}$ at roughly the same time. As transactions are selected for the next block, it is up to the miner to order the transactions submitted by $R_{1}, R_{2}$ and $R_{3}$ as long as the order of transactions submitted by a single user is preserved (e.g, given $R_{1}$'s transactions, $tx_{2}$ may not be executed before $tx_{1}$). Given that they all pay the same fee and there's no profit to be gained by choosing one over the other, the miner orders them arbitrarily such that those submitted by $R_{1}$ are executed first.
This causes the transactions delivered by $R_{1}$ to be executed successfully and leads those delivered by $R_{2}$ and $R_{3}$ to fail and ultimately be reverted, spending the fees they paid for transaction execution. This outcome not only causes $R_{1}$ to be the only relayer rewarded for their work but also incurs a loss for $R_{2}$ and $R_{3}$.

\item \textit{Scenario II -} $R_{1}, R_{2}$ and $R_{3}$ attempt to deliver $tx_{1}, tx_{2}, tx_{3}$ to $Chain_{B}$ as soon as they are included in $Chain_{A}$. This time, however, $R_{3}$ chooses to pay an increased gas fee for each one of the transactions while $R_{1}, R_{2}$ both pay a smaller, default fee. All transactions reach $Chain_{B}$ at roughly the same time. A rational miner chooses to include the transactions submitted by $R_{3}$ first as they are more profitable. This leads $R_{3}$ to be rewarded, albeit less given the amount required to cover the higher fees it paid for transaction submission. It also causes $R_{1}$ and $R_{2}$ to have their transactions reverted and their fees spent. In this case, $R_{3}$ managed to get its transaction executed first by choosing to forgo a portion of the rewards. It may also be possible that all relayers choose to pay increased fees, leading to even greater losses for those whose transactions are not executed first.

\item \textit{Scenario III -} When $tx_{1}, tx_{2}$ and $tx_{3}$ are included in $Chain_{A}$, both $R_{1}$ and $R_{2}$ attempt to deliver all of them to $Chain_{B}$. $R_{3}$, however, chooses to deliver only $tx_{1}$ and $ tx_{2}$ first, before attempting to deliver $tx_{3}$. $R_{1}$ and $R_{2}$ deliver the transactions at approximately the same time. $R_{3}$ however, spends less time retrieving transaction data from $Chain_{A}$ and processing it since it only attempts to deliver two out of the three available transactions. This leads $R_{3}$ to deliver $tx_{1}$ and $tx_{2}$ before block $h$, whereas both $R_{1}$ and $R_{2}$ deliver their transactions after block $h$ and before block $h + 1$. When block $h$ is minted, the miner includes the transactions $tx_{1}$ and $tx_{2}$ delivered by $R_{3}$. For the next block, the miner decides between $R_{1}$'s and $R_{2}$'s $tx_{3}$, based on the amount of fees they paid for the transaction. In this scenario, $R_{2}$ takes advantage of the fact that processing and delivering fewer transactions requires less time and claims the reward for both $tx_{1}$ and $tx_{2}$, leaving the other relayers to compete for the delivery of $tx_{3}$.

\end{itemize}

\textbf{Performance and scalability.} Performance issues exist as a direct consequence of the competition among independent relayers. Increasing the number of workers does not lead to an increase in cross-chain throughput.
This issue is compounded by the fact that, in the presence of multiple relayers, performance is decreased during times of high network volume~\cite{chervinski2023analyzing}. Moreover, when there is a surge in the number of pending transactions, relayers may attempt to retrieve and deliver a large amount of data at once. This may cause them to crash or behave unexpectedly~\cite{osmosislargeblock, websocketerror}. As a consequence, in a system where relayers work independently, not only performance is unable to scale but it may also be negatively impacted.

Another performance concern is derived on the reliance of relayers on blockchain full nodes to retrieve transaction data. For example, on Tendermint-based blockchains, the RPC server run by full nodes processes data queries sequentially. This may degrade performance and significantly increase waiting times when relayers attempt to retrieve data for a large amount of transactions at once~\cite{chervinski2023analyzing}. Distributing tasks among relayers can assist in mitigating this problem as relayers often run their own blockchain nodes for data retrieval purposes. In this manner, different nodes can process different subsets of the data queries concurrently.

\textbf{Transaction reordering and censorship.} Due to the open nature of distributed systems the set of relayers connecting a pair of blockchains is likely to be composed by heterogeneous machines with access to different network infrastructure. This  leads to some relayer operators being capable of processing and delivering cross-chain transactions faster than others. A similar situation happens in the presence of well-funded relayers who are willing to play a high-risk and high-reward game by paying increased transaction fees in an attempt to outbid its competitors. In either case, a single relayer or a group of colluding relayers may end up monopolizing the delivery of transactions between a pair of blockchains. In the long run, this might lead less powerful relayers to leave due to sustained financial loss. The remaining relayer or group of relayers will then be capable of selectively censoring transactions. 

In the absence of mechanisms to enforce cross-chain transaction ordering, monopolizing relayers can also reorder transactions to perform front-running and sandwich attacks~\cite{daian2020flash, heimbach2022eliminating}. 
For example, Tendermint-based blockchains, which are widely used for cross-chain communication together with the IBC Protocol, utilize a first-in first-out memory pool implementation where transactions are ordered based on the order in which they arrive in a node's memory pool. This can be exploited by a relayer with better hardware and infrastructure than its competitors. Similarly, in blockchains where miners are free to order transactions with the purpose of maximizing profit, a well-funded relayer can manipulate the order in which cross-chain transactions are executed by paying increased fees.

\subsection{Our proposal to improve cross-chain relaying}

Improving performance and eliminating the issues present in current cross-chain message relaying models requires enabling coordination between independent, potentially distrustful relayers. One way to achieve this is by designing a protocol to distribute the message delivery load among all available relayers, allowing each of them to work independently on a subset of the pending messages. However, merely distributing messages does not suffice for the purpose of achieving coordination in a permissionless setting. Relayers must also have an incentive to follow the protocol. While relayers might be motivated to cooperate and deliver only their respective subset of messages for the benefit of the network, they are also compelled to deviate from the protocol by the prospect of increasing their profit, e.g, by getting paid for delivering more messages. As a consequence, the aforementioned issues are likely to persist. Taking this into consideration, we conclude that in order for a coordination protocol to be effective in a permissionless setting it must implement adequate incentive mechanisms. In this light, we take the first step towards addressing the issues presented in this section by proposing a protocol to coordinate cross-chain message delivery, particularly in scenarios involving multiple relayers.

\section{System Model}\label{sec:system_model}
\subsection{Blockchains and transactions} 
We consider two independent, tamper-proof and append-only blockchains, $Chain_{A}$ and $Chain_{B}$. Both satisfy the persistence and liveness properties of distributed ledgers~\cite{garay2015bitcoin}. A blockchain consists of a sequence of blocks, each containing a set of transactions that modify the state of the system when executed. Transactions are submitted by users, participants who possess a blockchain account secured by a public and a private key. Blocks and their contents are publicly visible and are generated according to a protocol that is specific to each blockchain.

\subsection{Contracts} 
A contract is a publicly visible, deterministic program deployed in a blockchain. Each contract has an account capable of sending and receiving tokens, but unlike user accounts, they are controlled by contract code. Contracts define a set of functions that can be invoked by users to read, perform operations and write data to the ledger where it resides. Functions are invoked by submitting transactions containing function calls to the blockchain where the contract is deployed. Contracts cannot directly access information outside the ledger, but can receive external data (such as from other ledgers) as input.

\subsection{Network model}
We assume a synchronous network, where messages are delivered within a known upper bound of delay. Communication is reliable and messages are not lost after being sent.

\subsection{Cross-chain communication}
Blockchains are unable to directly exchange information with any external systems.
To communicate, two blockchains implement a common cross-chain messaging format such as XCM~\cite{xcmpolkadot}, and employ relayers to perform the transport of information between them. We refer to the connection between a pair of blockchains as a channel and say that relayers who work to deliver messages between $Chain_{A}$ and $Chain_{B}$ maintain the channel between those systems. When a blockchain sends information through the channel, it is referred to as the source. Conversely, when it receives information, it is referred to as the destination.

\subsection{Threat model} We consider relayers who attempt to maximize their profit, even if it such behavior seems irrational or requires them to deviate from the protocol. We do not consider attacks against the blockchains or contracts.

\subsection{Cryptographic assumptions}
Blockchain users have a public and private key. Public keys are used to receive transactions and are known to all system participants. Private keys are used to sign transactions and prevent forgery. Our protocol assumes a uniformly distributed, collision-resistant cryptographic hash function.

\section{Design goals}
Our objective is to design a protocol that enables coordination between multiple relayers working to deliver messages for a cross-chain channel. By achieving coordination we address two issues with current relaying models. First, we allow the cross-chain message delivery load to be balanced among available relayers, increasing the throughput of cross-chain communication. Second, we eliminate race conditions, encouraging more relayers to join the network and consequently reducing the chance of transaction censorship attacks. We now put forth three properties that are key to achieving our goals and use them to guide the design of our protocol. 

\textbf{Property I }(\textit{Scalability}). Scalability specifies that the throughput collectively achieved by the set of relayers that maintain a cross-chain channel should increase when the number of relayers increases. 

Because different relayers may have access to different hardware and network resources, performance cannot be guaranteed to be directly proportional to the number of relayers. Therefore we only assume that when the number of relayers maintaining a channel increases from $n$ to $n + 1$, the throughput of the channel also increases. By satisfying this requirement a protocol is able to adapt to a growth in cross-chain message delivery demands by employing more relayers.

\textbf{Property II }(\textit{Accountability}). Accountability specifies that a relayer must be held responsible, particularly through the imposition of punishment, should they fail to deliver messages assigned to them before their specified timeout. 

This property is essential for protocols operating in permissionless environments. Well-designed accountability mechanisms contribute to minimizing the occurrence of misbehavior and help ensure that participants act in the best interest of the network. 

\textbf{Property III }(\textit{Fair allocation}). This property specifies that a channel's message delivery workload should be uniformly distributed among the set of relayers maintaining the channel. 

We consider this strategy fair as it ensures that every relayer has an equal opportunity to deliver messages and profit from supporting the network. 
While this approach may not be ideal in the context of traditional distributed systems, where load balancing algorithms can tailor the distribution of tasks according to the performance of each specific machine, blockchains present unique constraints that make the implementation of complex load balancing algorithms particularly challenging. We discuss this further in Section \ref{sec:discussion}.
\section{A protocol for relayer coordination}\label{sec:protocol_presentation}
The first step towards achieving coordination among relayers is enabling them to communicate. Despite being unable to directly exchange information with each other, relayers share a connection with the blockchains at both ends of the channel. We leverage this in our design and propose an on-chain mechanism to manage the cross-chain channel between a pair of blockchains. We call this mechanism the \textsc{Coordinator}. 

The \textsc{Coordinator} must be implemented by both of the connected blockchains. 
In the interest of loosening requirements and allowing our protocol to be deployed in different blockchains, we assume that each instance of the \textsc{Coordinator} is implemented as a contract, although its functionality may be provided by a module in systems that support modular blockchain applications, such as Substrate and Cosmos SDK. 

Deploying the contracts requires the assistance of an external party. This process, however, does not introduce any trust assumptions because contract code is publicly visible and deterministic. Once deployed, both contracts can be inspected for correctness before being used. As the contracts are deployed in different blockchains, they are unable to directly verify operations executed by one another. To overcome this limitation, cross-chain protocols implement a mechanism to keep track of the counterparty blockchain's state and validate incoming data. In our protocol, we assume the contracts implement a light client~\cite{chatzigiannis2022lightclient} and require relayers to append updated block headers to incoming cross-chain messages.

To enable coordination and achieve \textit{scalability, accountability} and \textit{fair allocation} our protocol requires a \textsc{Coordinator} to implement the logic for membership management, task allocation, message delivery and incentivization. 

\subsection{Membership management}
Active relayers must be tracked in order to perform the allocation of tasks and distribute incentives. We achieve this by requiring relayers to undergo a registration procedure when joining a channel.

\textbf{Registration.} Relayers can start the registration process by publishing a transaction that invokes the \texttt{register()} function implemented by the \textsc{Coordinator}. This function receives the relayer's public key $pubkey$ and an amount of collateral defined by the contract. The collateral is held and slashed to punish the relayer in case of misbehavior, providing \textit{accountability}. The contract then generates a unique identifier $id$ for the newly registered relayer. The relayer's public key, $id$ and amount of deposited collateral are stored by the contract and it is included in the set of active relayers $\mathcal{R}$.

\textbf{Withdrawal.}
When a relayer no longer wishes to work towards maintaining a channel it publishes a blockchain transaction containing an invocation to the \texttt{withdraw()} function. This function receives the relayer's public key and starts an unbonding period, after which the relayer can reclaim its collateral. When the function is triggered, the requesting relayer is immediately removed from $\mathcal{R}$ to prevent it from receiving new message delivery tasks. Pending tasks allocated to this relayer before the start of the unbonding period must still be delivered. In the event a relayer fails to deliver any of the pending messages, it is punished proportionally by having its collateral slashed before it can be reclaimed. To ensure that this punishment cannot be evaded, the unbonding period must be set to a duration that exceeds the longest timeout associated with the relayer's pending tasks. For example, suppose that a relayer wishes to withdraw from a channel but still needs to deliver one pending message that times out at block $h$. The unbonding period must not end before block $h + 1$ and ideally should be made longer to accommodate possible delays in message delivery and confirmation.
Another possibility is for a relayer with a small amount of collateral remaining to attempt to avoid punishment by ceasing to deliver messages after sending a withdraw request. In this scenario, the penalty for failed deliveries may exceed the amount of locked collateral, allowing the relayer to get away without paying the full penalty. As a countermeasure, the contract can enforce a minimum amount of remaining collateral in order to continue receiving tasks. Once this amount is reached the start of the unbonding period is automatically triggered.

\subsection{Task allocation}
To eliminate concurrency and achieve \textit{scalability}, message delivery tasks must be distributed among the relayers that maintain a channel. We approach this challenge with the aforementioned \textit{Fair allocation} property in mind and propose two task allocation approaches. These are based on the concept of modular hashing, a technique that helps in uniformly distributing tasks among the relayers. For simplicity, we consider only the transfer of tokens between blockchains in our examples, however, the message handling logic can be applied to any cross-chain operation implemented by the contracts.

\textbf{Approach 1.} This approach allocates transactions to relayers when they are submitted via the \textsc{Coordinator}.  As a prerequisite, the programming language used to implement the \textsc{Coordinator} contract must offer a method for obtaining the transaction hash of transactions that invoke the contract, analogous to the \texttt{getTxId()} method available to Hyperledger Fabric chaincode [21]. Task allocation is performed when a cross-chain operation is successfully initiated through the \texttt{Coordinator}. Contracts are unable to modify transactions, therefore users that wish to initiate cross-chain operations must submit data that follows the format supported and enforced by the contracts. This can be achieved by using an external application to prepare and submit transactions.
Suppose a user submits a transaction invoking the \texttt{transfer()} function, which implements the logic to send tokens from an account in a source blockchain to another account in a destination blockchain. As input, it receives the sender's account, the recipient's account, the amount of tokens to be transferred and a timeout for the operation. Upon successfully validating the transaction data, the contract escrows the user's tokens in order to prevent them from being spent while the cross-chain transfer is pending. Then, it retrieves the hash of the transaction that invoked the \texttt{transfer()} function, $txHash$. The transaction hash is used to perform a modulo hashing operation $i = H(txHash) \mod m$, where $i$ is the resulting index, $H$ is a cryptographic hash function and $m$ is the size of the registered relayer set $\mathcal{R}$. The resulting index is used to access the position of a relayer in $\mathcal{R}$ and get its corresponding $id$. The \textsc{Coordinator} maps the $id$ to $txhash$ and stores that information together with the transaction's timeout to create a pending delivery task. As the new transaction is included in the blockchain, relayers are able to retrieve information regarding its allocation. 

\textbf{Approach 2.} This approach requires users to compute the allocation of transactions and submit it to the \textsc{Coordinator}. As a prerequisite, the \textsc{Coordinator} contract must be capable of retrieving a recorded transaction for a given hash value. An example is the \texttt{getHistoryForKey()} method supported by Hyperledger Fabric chaincode~\cite{gettxidmethod}. Alternatively, the same logic may be implemented by utilizing an oracle service to query the blockchain for confirmed transactions. 

Similar to Approach 1, a user invokes a contract function to initiate a cross-chain operation and provides transaction data. The contract validates the transaction, but we assume that it does not have access to the hash of the contract invocation transaction, therefore it cannot perform the modulo hashing operation at this time. Instead, the contract implements an \texttt{assignTasks()} function that can be invoked to provide the contract with the allocation for the newly added transactions. This function receives a list of transaction hashes together with the $id$ of the assigned relayer. A user or relayer can perform task allocation by watching the blockchain for new cross-chain transactions and querying the \textsc{Coordinator} for the current set of registered relayers. This information can then be used to perform the modulo hashing operation. 
When the contract receives task allocation information via the \texttt{assignTasks()} function, it verifies if the received transaction hashes identify a cross-chain operation included in the blockchain and not yet assigned. If so, it executes the task allocation mechanism to check if the allocation was properly performed according to the set of registered relayers. In case the allocation is correct, information regarding the allocated tasks is stored by the contract and the transaction is published in the blockchain. Otherwise, the \textsc{Coordinator} marks the transaction as reverted. Users can be incentivized to perform task allocation by providing a reward for the first correct submission.

This approach incurs a drawback proportional to the blockchain's block interval time. Assume a blockchain in which the average block interval is equal to 10 seconds. Once transactions are included in the blockchain, users must perform the allocation and invoke the \texttt{assignTasks()} function. However, the function will only be executed when a new block is created, leading to a delay of at least 10 seconds before the task allocations are published in the blockchain. This delay will be proportional to the blockchain's block interval, therefore this approach needs to be considered in a case-by-case basis. We argue that for blockchains with a block interval of a few seconds, this trade-off is beneficial. According to recent work, when a channel is being maintained by a single relayer the latency for cross-chain transaction completion may be in the order of several minutes during times of high network load~\cite{chervinski2023analyzing}. Consequently, this approach trades a small delay, which is required to perform the distribution of the delivery load among relayers, for the ability to prevent prolonged wait times during periods of high network usage.

\subsection{Message delivery}
Relayers retrieve and deliver data by interacting with blockchain full nodes. To receive transaction delivery tasks from a blockchain's \textsc{coordinator}, relayers must be registered with it. To deliver a transaction, however, registration with the destination blockchain's \textsc{Coordinator} is not required. We discuss this design choice in detail in Section \ref{subsec:incentives}.

\textbf{Transaction flow.} In our model we consider that cross-chain operations are composed by three steps, each requiring one transaction to be included in the communicating blockchains. The first transaction is included in the source blockchain to \textit{request} a cross-chain operation, the second transaction is included in the destination blockchain as a \textit{receipt} for the processed the operation, the third and final transaction is included in the source blockchain as an \textit{acknowledgement} of the receipt and the result of the operation. When a relayer is assigned to a \textit{request} transaction it becomes responsible for retrieving it from the source, submitting it to the destination and bringing the receipt back to the source blockchain.

When registered, a relayer continually scans the blockchain for new cross-chain transactions and their allocation information. Upon encountering a task assigned to itself, it retrieves the corresponding transaction data from the source blockchain, formats it according to the appropriate cross-chain logic (e.g, by adding extra information) and submits it by invoking the \texttt{deliverTx()} function implemented by the \textsc{Coordinator} in the destination blockchain. This function receives transaction data and validates it according to the cross-chain communication logic implemented by the \textsc{Coordinator}. The contract then executes the logic required by cross-chain operation in the destination blockchain. For a token transfer, for example, it may mint new tokens and send them to the recipient account specified in the received transaction data. The transaction is then included in the destination blockchain. The relayer retrieves data corresponding to the \textit{receive} transaction from the destination blockchain, performs additional formatting if required and submits a transaction to the source blockchain to invoke the \texttt{proveDelivery()} function implemented by the \textsc{Coordinator}. This function receives the transaction receipt included in the destination and validates it. The contract may then execute additional logic to complete the operation, if required. In the case of a token transfer, the contract may destroy tokens previously held by the contract when it was initiated. After the received information is processed, the transaction is included in the blockchain to notify the \textit{acknowledgement} of the operation.

\textbf{Handling timeouts.} Transactions may time out, either due to congestion in the destination blockchain or due to relayer inactivity. In those cases, the timeout may be informed to the source blockchain by invoking the \texttt{submitTimeout()} function. This function receives and verifies a proof-of-absence, which demonstrates that a given transaction was not included in the destination blockchain within the expected block range. Examples of how such proofs may be constructed can be found in existing work~\cite{yu2017decim, yu2016dtki}.

\subsection{Relayer incentives}\label{subsec:incentives}
Providing adequate incentives is essential to ensure that participants will follow the protocol. We address this challenge by designing a mechanism that provides \textit{accountability} by rewarding successful transaction deliveries, punishing inactivity and discouraging relayers from stealing allocated transactions. As a consequence the proposed mechanism also assists in achieving \textit{scalability} by encouraging relayers to work on separate delivery tasks concurrently.

\textbf{Delivery fees.} Users incentivize relayers through the payment of a delivery fee for completed tasks. The corresponding tokens are sent to the \textsc{Coordinator} and held when a cross-chain transfer is initiated. 
To encourage relayers to take responsibility over all the data delivery steps required for the completion of a cross-chain operation, delivery fees should only be released in the source blockchain when the delivery receipt is brought from the destination. In addition, the reward should always be sent to the relayer to which the task was initially allocated, regardless of who delivered it. This strategy achieves two important goals. First, it compels relayers to deliver a \textit{request} from the source to the destination, retrieve the corresponding \textit{receipt} and submit it back to the source as an \textit{acknowledgement} of the operation before they can earn rewards. Second, it prevents a relayer from earning rewards by delivering transactions allocated to a different relayer. Deviating from the protocol by abandoning an ongoing operation or stealing transactions yields no rewards. This is compounded by the fact that submitting transactions to invoke contract functions requires the payment of gas fees. Consequently, if relayers deviate from the protocol they sacrifice their rewards and incur a loss due to transaction submission fees.

Delivery fees should be carefully considered in order to guarantee the generation of profit for relayers. Considering the two transactions that must be submitted to complete an operation (\textit{receipt} and \textit{acknowledgement}), the corresponding reward should exceed the fees paid to submit those transactions to the destination and source blockchains, respectively.

\textbf{Punishing misbehavior.} When a transaction times out, the relayer responsible for its delivery is punished by having its collateral slashed. This strategy, however, does not encourage a relayer to report timeouts as it leads to a loss of funds. In order to guarantee that timeouts are reported back to the source blockchain, other relayers may be rewarded for submitting a proof to report the timeout of a transaction to the source blockchain's \textsc{Coordinator}. They may track the timeouts by comparing the transaction allocation information stored in the source blockchain's \textsc{Coordinator} contract against the state of the destination blockchain. The first submission of a valid proof of timeout is rewarded with tokens deducted from the inactive relayer's collateral. 

When a transaction times out, the user that initiated the operation also incurs a loss due to the gas fees spent on transaction submission. In order to mitigate this, users can also receive a portion of the tokens deducted from the inactive relayer's collateral to cover the cost of their transaction.

\section{Discussion and research challenges}\label{sec:discussion}
Similar to classical distributed systems, there is no one-size-fits-all approach when it comes to balancing workloads and allocating tasks in the context of cross-chain communication. The specific requirements of each application and the environment it will be deployed in must be considered to pick a suitable strategy. This often means prioritizing some features over others, resulting in trade-offs. We now discuss those trade-offs and design choices.

\textbf{Performance vs. reliability.}
By enabling relayers to work on independent tasks concurrently, a protocol achieves scalability and becomes capable of adapting to increases in cross-chain transaction volume. Although this improves performance, it removes redundancy and consequently reduces the reliability of a relaying service. When different relayers attempt to deliver the same task simultaneously, the likelihood of it being successfully delivered is higher compared to when it is assigned to a single relayer. When tasks are allocated with the objective of eliminating redundant work, system liveness may be compromised if several relayers stop responding simultaneously. In contrast, in current models, as long as a single relayer remains operational liveness is maintained (assuming that it can handle the channel load). A potential solution to this problem is an incentive mechanism that discourages relayers from stealing other's tasks but incentivizes their delivery if they are near their timeout. Other possibilities are allowing redundancy to be configured through a parameter that dictates the number of relayers which will receive the same task. A system may also adapt to demand dynamically, prioritising performance over redundancy as the volume of transactions increases. In such systems, incentive models should be considered carefully to prevent inactive relayers from being rewarded for the effort of others, a phenomenon known as free-riding.

\textbf{Efficient task allocation.} We acknowledge that while the simplicity of our task allocation mechanisms make them easier to understand and implement, it may also be seen as a limitation of our work.
Distributed systems are often composed by participants with heterogeneous resources. For this reason task allocation can be improved by tailoring each relayer's workload to maximize performance without exceeding their capabilities as it may lead to delays or crashes. While simple task allocation strategies are easier to implement and demand less resources to be executed, their performance may be inferior to more sophisticated, resource-intensive alternatives. Finding the optimal point between resource consumption and performance is a challenge when designing task allocation strategies for blockchain-based systems and is complicated further due to their unique constraints. For example, blockchains proceed in block intervals that may range from a few seconds to several minutes in different systems. Due to this, any change to previously assigned jobs, such as migrating it to another relayer, can only be published to the blockchain upon the creation of a new block and may incur a long delay. Another constraint is the inability of blockchains to access information regarding the hardware and infrastructure resources of participants. This can be addressed by dynamically computing the performance of each individual relayer as it completes tasks and using the resulting metrics as a criterion for task allocation. This, however, can be prohibitively costly if allocation is managed by smart contracts, given the substantial amounts of fees they require to perform complex computations.

\textbf{Incentives and quality of service.} If given the choice, honest, rational relayers will always prioritize the delivery of tasks that yield more profits. For this reason, users often choose to pay higher delivery fees when they wish their transactions to be prioritized. When tasks are automatically allocated relayers are denied the opportunity to select them based on the corresponding reward and users become unable to provide extra incentives to give their transaction priority. Another consequence of removing the reward dynamics in the context of message relaying is the potential impact in the urgency with which relayers deliver tasks. In a competitive environment, relayers are motivated to invest in better infrastructure in order to maximize their chance of profit, leading to an increase in the quality of the service provided. Conversely, in the absence of competition, relayers are not encouraged to invest resources in order to maximize their output. Therefore, while winner-takes-all games pose a disadvantage to less powerful relayers, they simultaneously encourage participants to provide better service in order to earn profits and as a consequence benefit the system.

\section{Related Work}\label{sec:related_work}
Despite an abundance of works concerned with strategies to achieve communication between independent blockchains, few explore the performance of cross-chain communication~\cite{belchior2023hephaestus, mihaiu2021framework} and in particular, data relaying systems~\cite{chervinski2023analyzing, olshansky2023relay}.  To the best of our knowledge, we are the first to identify the issues associated with permissionless cross-chain relaying systems. We are also the first to take action towards addressing those issues by proposing a protocol to perform task allocation and achieve coordination among independent relayers.

The subject of task allocation and load balancing in classical distributed systems has been widely studied, with works dating from several decades ago~\cite{srinivasan1999safety,jiang2015survey, bannister1983task}.
Such strategies have also been investigated in systems composed by non cooperative participants~\cite{grosu2005noncooperative, kameda2012optimal, roughgarden2001stackelberg}. Those subjects, however, have not been explored in the context of cross-chain communication.

\section{Conclusion}\label{sec:conclusion}
Relaying services play an important role in the transport of data between blockchains, however, their performance and shortcomings have not been thoroughly studied. In this work we identified issues that have a negative impact on cross-chain relaying services, including race conditions, lack of scalability and security concerns. We took the first step towards addressing some of those issues by proposing a protocol to coordinate independent cross-chain relayers in a permissionless setting. Additionally, we leveraged our protocol as a foundation for a discussion about the trade-offs that require consideration when designing relayer coordination protocols. We conclude that further research and development are required in cross-chain transaction ordering, cross-chain performance analysis and load balancing among independent cross-chain relayers.

\bibliographystyle{IEEEtran}
\bibliography{references}

\end{document}